\documentclass[11pt,twoside]{article}
\usepackage{asp2010}
\usepackage{graphicx}
\resetcounters

\begin{document}

\title{Fast Rotating solar-like stars using asteroseismic datasets}
\author{R.~A. Garc\'\i a $^{*,1}$, T. Ceillier$^1$, T. Campante$^2$, G.~R. Davies$^1$, S.~Mathur$^{3}$, J.~C~Su\'arez$^4$, J. Ballot$^{5,6}$, O. Benomar$^7$, A. Bonanno$^8$, A.~S.~Brun$^1$, W.~J.~Chaplin$^9$,  J.~Christensen-Dalsgaard$^2$, S. Deheuvels$^{10}$,Y.~Elsworth$^9$, R.~Handberg$^2$, S.~Hekker$^{11}$, A.~Jim\'enez$^{12,13}$, C.~Karoff$^2$, H.~Kjeldsen$^2$, S.~Mathis$^1$, B.~Mosser$^{14}$, P.~L.~Pall\'e$^{12,13}$, M.~Pinsonneault$^{15}$, C.~R\'egulo$^{12,13}$, D.~Salabert$^{16}$, V.~Silva~Aguirre$^{17}$, D.~Stello$^7$, M.~J. Thompson$^{3}$, G. Verner$^9$, and the PE11 team of {\it Kepler}  WG\#1\\
{$^*$} Affiliations are given at the end of the paper}
%\affil{$^1$Institution Full Address for Author1}
%\affil{$^2$Institution Full Address for Author2}
%\affil{$^3$Institution Full Address for Author3}
%}

\begin{abstract}
The NASA {\it Kepler mission} is providing an unprecedented set of asteroseismic data. In particular, short-cadence lightcurves ($\sim$~60s samplings), allow us to study solar-like stars covering a wide range of masses, spectral types and evolutionary stages. Oscillations have been observed in around 600 out of 2000 stars observed for one month during the survey phase of the {\it Kepler} mission. The measured light curves can present features related to the surface magnetic activity (starspots) and, thus we are able to obtain a good estimation of the surface (differential) rotation.  In this work we establish the basis of such research and we show a potential method to  find stars with fast surface rotations.
\end{abstract}

\section{Introduction}
Solar-type stars have global physical characteristics similar to the Sun. From the point
of view of their oscillations, solar-like stars are excited by turbulent motions in their convective outer layers \citep[e.g.][]{1977ApJ...212..243G,2008A&A...478..163B}. These oscillations are also present in red-giant stars \citep[e.g.][]{2009Natur.459..398D,2010ApJ...713L.176B} and even in massive B-type stars as recently shown by \citet{2009Sci...324.1540B}.

Solar-type stars are generally slow rotators (in most cases $v \sin i<20\,$km s$^{-1}$) and the influence of rotation on the oscillation frequencies is currently considered to be small (at least) on a global scale. However, distortion due to the centrifugal force can have a larger impact on the oscillation frequencies even for slow rotators \citep[][ and references there in]{2009LNP...765...45G,2010AN....331.1038R}. Such an effect is stronger for p modes with small inertia. These modes propagate mainly through the outer layers of the star. Therefore, their frequencies are more sensitive to changes in the surface physical properties, where the centrifugal force becomes more efficient \citep[e.g.][]{2010ApJ...721..537S}. Moreover, when the rotation becomes important, the coriolis acceleration influences the stochastic excitation of the modes and the amplitudes at the surface could be different for the $\pm$~$m$ components of a given multiplet \citep{2009A&A...508..345B}. Finally, the structure and evolution of the stars change when rotation is introduced into the structure and evolution models \citep[e.g.][]{2004A&A...425..229M,2009A&A...495..271D,2010A&A...519A.116E}.
Once these effects are taken into account the inferred age of the star could change, which has severe consequences, for example, to the exoplanet research.

Space facilities such as CoRoT \citep{2006cosp...36.3749B} and {\it Kepler}  \citep{2010Sci...327..977B}, allow us to measure oscillations in many solar-like stars \citep[e.g.][]{2008Sci...322..558M,2011Sci...332..213C}, and thus to potentially study their rotation. Rotation breaks the degeneracy in the azimuthal order $m$ of the modes, by splitting the modes in $2 \ell + 1$ components. Depending on the inclination angle of the star only a given number of these components are visible  \citep[e.g.][]{2003ApJ...589.1009G,BalGar2006}. However, when the damping time of the modes is small, the modes have large widths in the frequency domain and the split $m$ components are overlapped, making the measure of the splitting and thus of the internal rotation difficult. This effect is particularly important in F-type stars \citep[e.g.][]{2008A&A...488..705A,2009A&A...506...51B,2009arXiv0907.0608G,2011ApJ...733...95M}. 

High-precision photometry, as needed by asteroseismology, also provides a measurement of the surface rotation rate through the detailed analyses of the light curves. When a star is active, starspots cross the visible stellar disk producing a reduction in the luminosity of the star that can be measured (see Fig.~\ref{LC_Sun}). The time evolution of such perturbations provides a measurement of the surface velocity at the latitudes of the spot, which also allows us to determine the surface differential rotation \citep[e.g.][]{2009arXiv0908.2244M,2010A&A...518A..53M}.

In the present work, we use the Sun-as-a-star measurements collected by the Solar PhotoMeters (SPM) of the Variability of solar IRradiance and Gravity Oscillations (VIRGO) instrument \citep{1995SoPh..162..101F} on board the Solar and Heliospheric Observatory (SoHO) in order to help guide the analysis of the rotation on other stars. In section 2 we explain the methodology and we apply it to the Sun. In section 3 we will show some preliminary results obtained using the Survey phase of the {\it Kepler} satellite.

\section{Methodology}

To study the surface rotation rate of the asteroseismic targets, we analyse, in a semi-automatic way, the low-frequency part of the power spectrum. We concentrate on the region with periods longer than one day with a lower limit corresponding to 1/4 of the total length of the time series considered.  We look for peaks above a given threshold whose value is computed in a statistical way after analyzing more than 2000 solar-like stars measured during the {\it Kepler} survey phase \citep[e.g.][]{2011MNRAS.tmp..892V}.  A starspot fitting can then be performed in the most promising stars. This technique can give more information about the properties of the main spots as well as a detailed description of the differential rotation in the active latitudes \citep[e.g.][]{2010A&A...518A..53M,2011A&A...530A..97B}. However it is very time consuming and it is not very well suited to analyse a very large number of stars. 

It is important to notice that the analysis based on the light curves, will not extract the surface rotation for all stars. Indeed, only those stars exhibiting starspots produce a measurable signature of the rotation. Moreover, if the inclination angle is small and the active latitudes of the star are close to the equator it is difficult to measure a signature of the rotation. This could be the case for the CoRoT target HD49385 \citep[for more details see:][]{2010A&A...515A..87D}.

\begin{figure}[!htbp]
\begin{center}
\includegraphics[width=8.2cm, angle=-90]{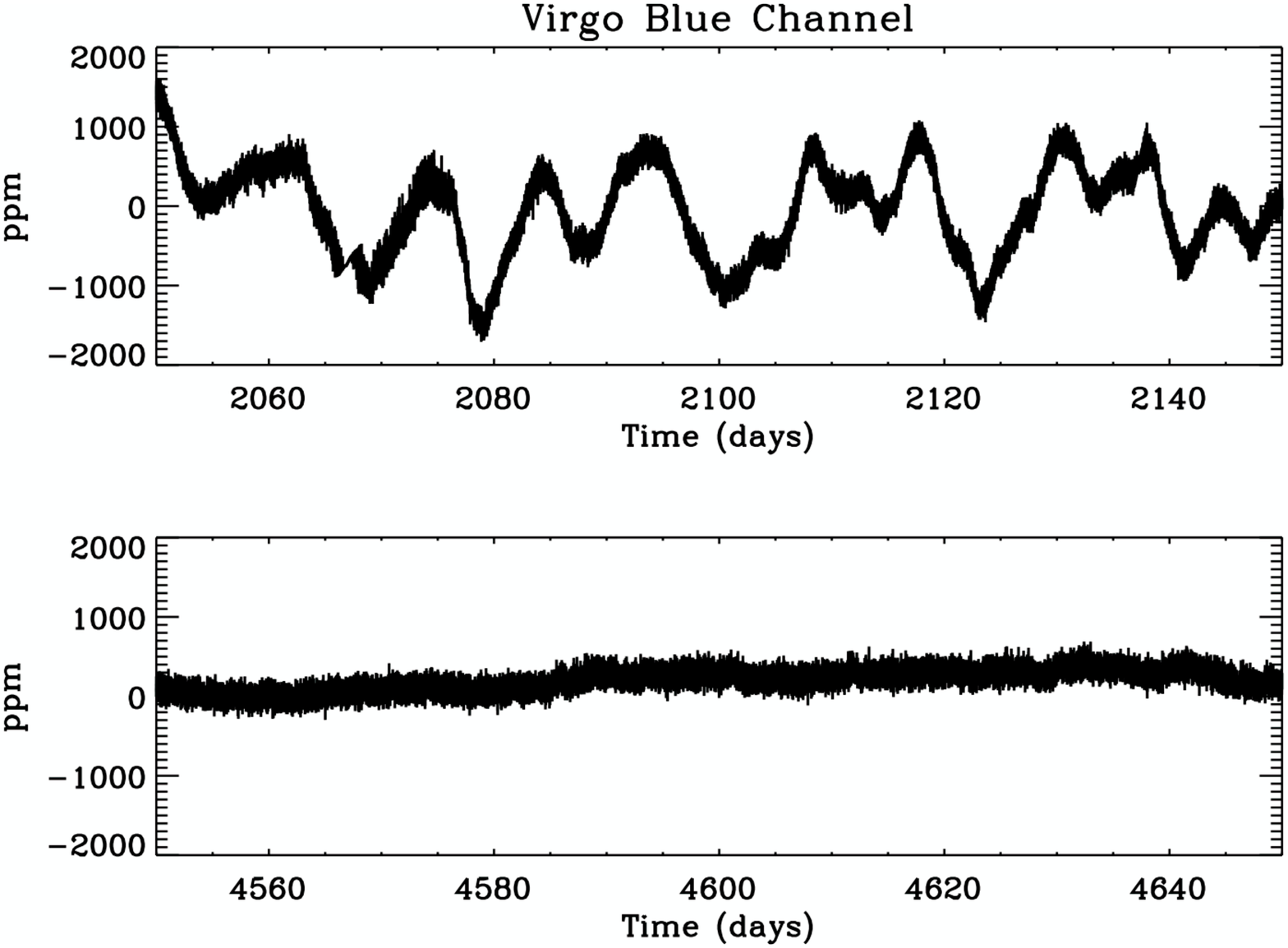}
\caption{Solar photometric light curves taken by the VIRGO/SPM instrument (blue channel) during a period close to the maximum  of the activity cycle 23 (top) and during the minimum between cycles 23 and 24 (bottom). The origin of time is March, 2, 1996.}
\label{LC_Sun}
\end{center}
\end{figure}

\begin{figure}[!htbp]
\begin{center}
\includegraphics[width=8.5cm, angle=90]{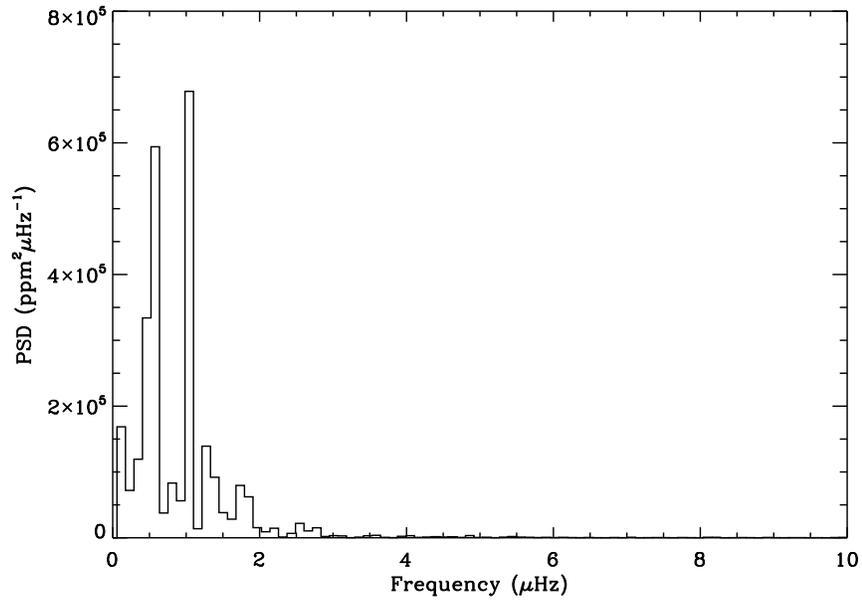}
\caption{Low-frequency region of VIRGO/SPM blue channel PSD obtained during 100 days close to the maximum of solar activity cycle 23.}
\label{fft_max}
\end{center}
\end{figure}

\begin{figure}[!htbp]
\begin{center}
\includegraphics[width=8.5cm, angle=90]{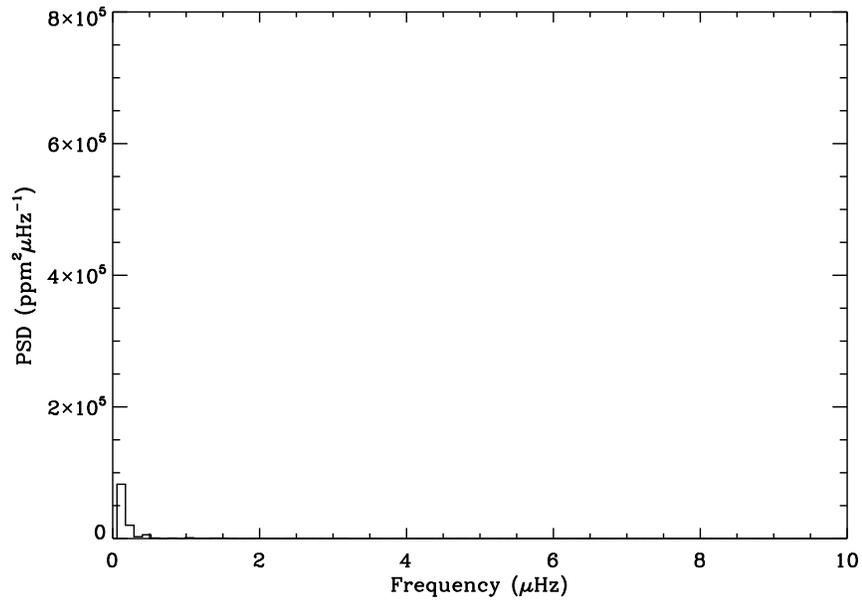}
\caption{Same legend as Fig.~\ref{fft_max} but using 100 days during the minimum of the activity between cycles 23 and 24.}
\label{fft_min}
\end{center}
\end{figure}

In Fig.~\ref{LC_Sun} we show 100 days of the VIRGO/SPM light curve taken close to the maximum of the activity cycle 23 (top) and 
during the extended minimum \citep{2009A&A...504L...1S} between cycles 23 and 24 (bottom). In the first case, the sunspots crossing the visible disk of the Sun produce a modulation of the flux in the light curve. In the second case, the light curve is flat and we do not see any suggestions of the presence of spots in the Sun. This is corroborated by the detailed analysis of the power spectral density (PSD). In Fig.~\ref{fft_max} and \ref{fft_min} we show the PSDs of the light curves taken at the maximum an during the minimum of solar activity. During the maximum, we can clearly distinguish a peak structure with a maximum around 0.9 $\mu$Hz and a secondary peak at $\sim$ 0.45 $\mu$Hz corresponding to $\sim$ 13 and $\sim$ 26 days respectively. In this example, the highest peak is the second harmonic and not  the first one. Therefore, when analyzing other stars, we need to be very careful and, when a peak is selected in the PSD, we need to check if there is a harmonic at lower frequency that could be the frequency associated with the rotation period we are looking for. Contrary to the PSD of the light curve taken during the maximum activity, at minimum, we do not see any peak associated to the rotation  (Fig.~\ref{fft_min}). 

\section{Analysis of asteroseismic light curves}

\begin{figure}[!htbp]
\begin{center}
\includegraphics[width=8.8cm, angle=90]{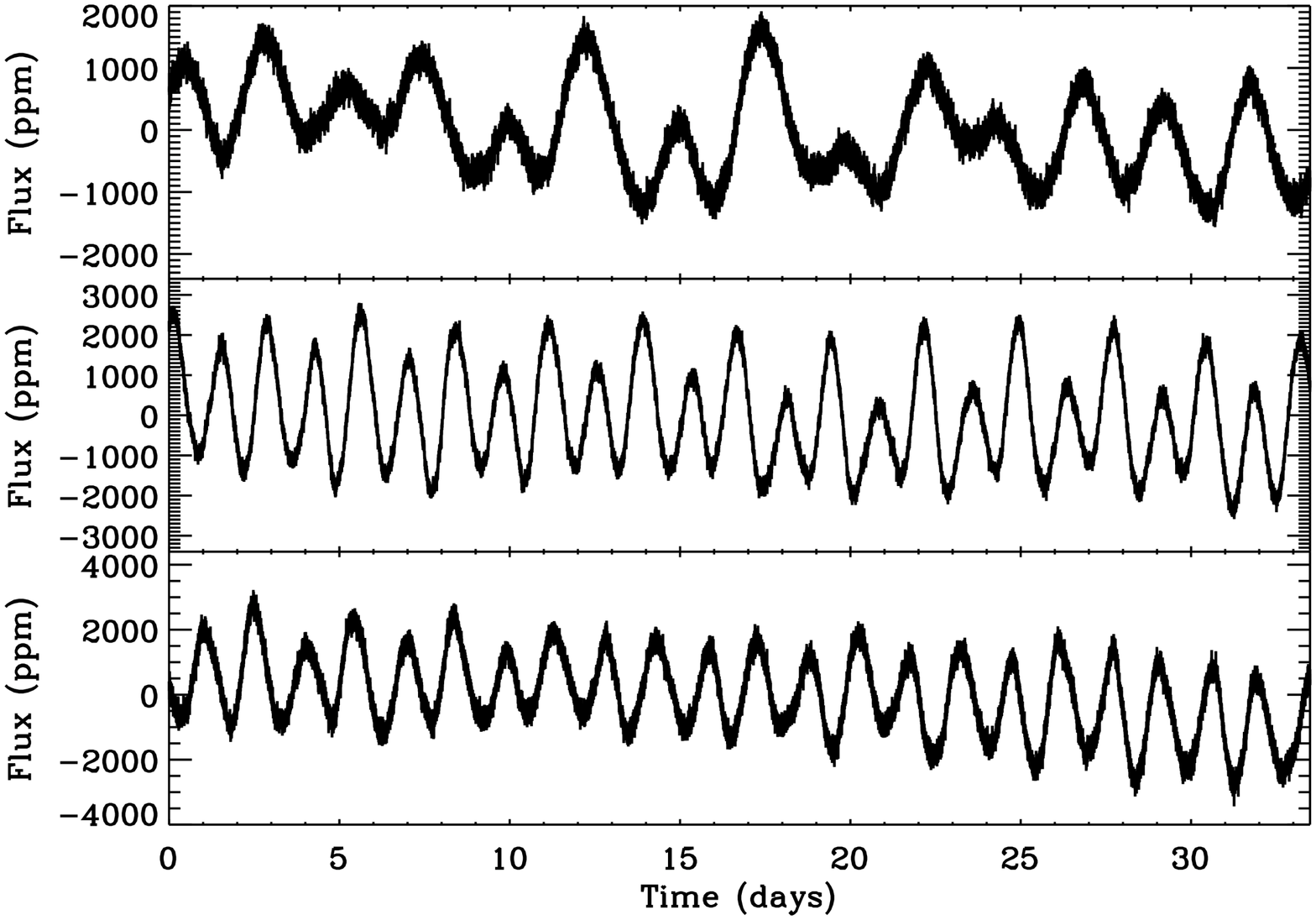}
\caption{Example of {\it Kepler} light curves showing starspots in their surface. Time starts on May 13, 2009.}
\label{fig_kep}
\end{center}
\end{figure}

Preliminary analyses of the {\it Kepler} data \citep[following the methods described in][]{2011MNRAS.414L...6G} show that many stars have signs of magnetic activity using a starspots proxy \citep[e.g.][]{2010Sci...329.1032G}. However, the amplitude of the oscillation modes and the activity proxy are anticorrelated. Unfortunatelly, those stars exhibiting higher magnetic activity have no detectable modes \citep{2011ApJ...732L...5C} in the PSD. In Fig.~\ref{fig_kep} we show the light curves of three typical stars in which some surface magnetic activity has been measured. A repetitive pattern of a few days can be seen in all of them. They are potentially fast surface rotators. More complete analyses, now being carried out by the PE11 team inside the {\it Kepler} working group 1, include the use of wavelet tools to obtain reliable periodicities for all the stars in which pulsations have been detected during the first year of measurements.

\acknowledgements 
Funding for the {\it Kepler} Discovery mission is provided by NA\-SAs Science Mission Directorate. The authors wish to thank the entire Kepler team, without whom these results would not be possible. We also thank all funding councils and agencies that have supported the activities of KASC Working Group 1, and the International Space Science Institute (ISSI). SoHO is an international collaboration program of ESA and NASA. VIRGO is the result of a cooperative effort of many individual scientists and engineers at several institutes in Europe and the USA to whom we are deeply indebted. NCAR is supported by the National Science Foundation. SH also acknowledges financial support from the Netherlands Organization for Scientific Research (NWO). RAG thanks the support of the French PNPS. \\
Affiliations: 
{$^1$La\-bo\-ra\-toi\-re AIM, CEA/DSM -- CNRS - Universit\'e Paris Diderot -- IR\-FU\-/SAp, 91191 Gif-sur-Yvette Cedex, France.}
{$^2$Danish AsteroSeismology Centre, Department of Physics and Astronomy, 8000 Aarhus C, Denmark.}
{$^3$High Altitude Observatory, NCAR, P.O. Box 3000, Boulder, CO 80307, USA.}
{$^4$Instituto de Astrof\'\i sica de Andaluc\'\i a (CSIC), Granada, Spain.}
{$^5$Institut de Recherche en Astrophysique et Plan\'etologie, Universit\'e de Toulouse, CNRS, 14 avenue E. Belin, 31400 Toulouse, France.}
{$^6$Universit\'e de Toulouse, UPS-OMP, IRAP, 31400 Toulouse, France.}
{$^7$Sydney Institute for Astronomy, School of Physics, University of Sydney, NSW 2006, Australia.}
{$^8$INAF Osservatorio Astrofisico di Catania, Via S. Sofia 78, 95123, Catania, Italy.}
{$^9$School of Physics and Astronomy, University of Birmingham, Edgbaston, Birmingham B15 2TT, UK.}
{$^{10}$ Department of Astronomy, Yale University, P.O. Box 208101, New Haven, CT 06520-8101, USA.}
{$^{11}$Astro\-no\-mi\-cal Institute ``Anton Pannekoek'', University of Amsterdam, PO Box 94249, 1090 GE Amsterdam, The Netherlands.}
{$^{12}$Uni\-ver\-si\-dad de La Laguna, Dpto de Astrof\'isica, 38206, Tenerife, Spain.}
{$^{13}$Ins\-ti\-tu\-to de Astrof\'\i sica de Canarias, 38205, La Laguna, Tenerife, Spain.}
{$^{14}$LESIA, UMR8109, Universit\'e Pierre et Marie Curie, Universit\'e Denis Diderot, Obs. de Paris, 92195 Meudon Cedex, France.}
{$^{15}$Astronomy Department, Ohio State University, Columbus, Ohio 43210, USA.}
{$^{16}$Universit\'e de Nice Sophia-Antipolis, CNRS, Observatoire de la C\^ote dÕAzur, BP 4229, 06304 Nice Cedex 4, France.}
{$^{17}$Max Planck Institute for Astrophysics, Karl-Schwarzschild-Str. 1, 85748, Garching bei M\"unchen, Germany.}

%{$^3$Centro de Astrof\'isica and Faculdade de Ci\^encias, Universidade do Porto, Rua das Estrelas, 4150-762 Porto, Portugal.}
%{$^5$Institut d'Astrophysique Spatiale, UMR8617, Universit\'e Paris XI, Batiment 121, 91405 Orsay Cedex, France.}
%{$^{12}$Astro\-no\-mi\-cal Institute ``Anton Pannekoek'', University of Amsterdam, PO Box 94249, 1090 GE Amsterdam, The Netherlands.}
%{$^{16}$Australian Astronomical Observatory, PO Box 296, Epping NSW 1710, Australia.}
%{$^{17}$Departamento de F\'isica e Astronomia, Faculdade de Ci\^encias da Universidade do Porto, Portugal.}
%{$^{18}$Institute of Astronomy, University of Vienna, A-1180, Vienna, Austria.}
%{$^{19}$Astronomical Institute, University of Wroc{\l}aw, ul. Kopernika 11, 51-622 Wroc{\l}aw, Poland.}
%{$^{20}$Space Telescope Science Institute, Baltimore, MD 21218, USA.}
%{$^{21}$Department of Physics and Astronomy, Iowa State University, Ames, IA 50011, USA.}
%{$^{22}$SETI Institute/NASA Ames Research Center, Moffett Field, CA 94035, USA.}
%{$^{23}$NASA Ames Research Center, Moffett Field, CA 94035, USA.}

\bibliographystyle{asp2010} 
\bibliography{./BIBLIO}

%\bibliography{aspauthor}

\end{document}